\documentstyle[fleqn,iopfts,12pt]{ioplppt}
\jl{1}
\eqnobysec
\begin{document}
\title{On the uncertainty relations and squeezed states for the
quantum mechanics on a circle}
\author{K Kowalski and J Rembieli\'nski}
\address{Department of Theoretical Physics, University
of \L\'od\'z, ul.\ Pomorska 149/153,\\ 90-236 \L\'od\'z,
Poland}
\begin{abstract}
The uncertainty relations for the position and momentum of a quantum particle
on a circle are identified minimized by the corresponding coherent
states.  The sqeezed states in the case of the circular motion are  introduced
and discussed in the context of the uncertainty relations.
\end{abstract}
\newpage
\section{Introduction}
The uncertainty relations are one of the most fundamental concepts
of quantum theory.  In spite of its importance and long history [1--8]
the problem of finding such relations in the case of the quantum mechanics 
on a circle still remains open.  In fact, the experience with the
standard Heisenberg uncertainty relations suggests that the
uncertainty relations for the quantum mechanics on a circle should
be related with the corresponding coherent states.  Nevertheless,
the existing approaches connecting the uncertainty relations with
the coherent states can hardly be called satisfactory.

In this work we introduce the new uncertainties for the position and
momentum of a quantum particle on a circle and new uncertainty
relations referring to the very recently found coherent states for 
the circular motion [9,10].  We also introduce the squeezed states for 
the quantum mechanics on a circle and discuss them in the context of
the uncertainty relations.

We begin with a brief account of the alternative approaches linking
the uncertainty relations for a quantum particle on a circle to
the coherent states.  As far as we are aware there are only two such
approaches.  In the first one we deal with the uncertainty relations
implied by the $e(2)$ algebra satisfied by the angular momentum
operator and the cosine and sine of the angle operator 
\begin{equation}
%<1.1>
[\hat J, \cos\hat\varphi]={\rm i}\hbar\sin\hat\varphi,\qquad
[\hat J, \sin\hat\varphi]=-{\rm i}\hbar\cos\hat\varphi,\qquad 
[\sin\hat\varphi,\cos\hat\varphi]=0.
\end{equation}
These relations are of the form
\numparts
\begin{eqnarray}
%<1.2>
\Delta \hat J\Delta \cos\hat\varphi &\ge& \frac{\hbar}{2}|\langle
\sin\hat\varphi\rangle|,\\
\Delta \hat J\Delta \sin\hat\varphi &\ge& \frac{\hbar}{2}|\langle
\cos\hat\varphi\rangle|,\\
\Delta \sin\hat\varphi\Delta\cos\hat\varphi &\ge& 0.
\end{eqnarray}
\endnumparts
The states minimizing (1.2{\em b}) [3] are referred to as the circular
squeezed states.  Recently, those states have been applied in the
study of the Rydberg wave packets [8].  We point out that (1.2{\em a}) and
(1.2{\em b}) cannot be minimized simultaneously [8].  Let us write down
the (normalized) wave packets corresponding to the circular squeezed
states, i.e., the position representation of these states in the
space of square integrable functions on a circle $L^2(S^1)$.  We have
\begin{equation}
%<1.3>
f_{\alpha,l}(\varphi) =
\frac{1}{\sqrt{2\pi I_0(2s)}}\exp[s\cos(\varphi -\alpha)+{\rm
i}l(\varphi-\alpha)],
\end{equation}
where the packet is peaked at $\alpha$, $l=\langle \hat J\rangle$ is
the expectation value of the angular momentum, $s$ is the squeezing
and $I_0$ is a modified Bessel function of the first kind.  Of
course, the wave packet on a circle should be $2\pi$-periodic.  In
view of (1.3) this implies that $l$ is integer.   But the classical
angular momentum is an arbitrary real number.  Therefore, the
circular squeezed states are not labelled by points of the classical
phase space.  Bearing in mind that the standard coherent states for
a particle on a real line are marked with points of the classical
phase space we conclude that (1.3) are rather poor candidate to
represent the coherent states for a particle on a circle.  On the
other hand, it turns out that the uncertainty relations (1.2) cannot
be used for determining the correct coherent states for the quantum
mechanics on a circle.  In our opinion the genuine coherent states
for a quantum particle on a circle are those introduced in our joint
paper [9] as a solution of some eigenvalue equation (see the next
section), and, independently, by Gonz\'ales and del Olmo [10] who applied 
the Weil-Brezin-Zak transform.  An attempt to connect these coherent
states for the quantum mechanics on a circle with the uncertainty
relation of the form [6,10]
\begin{equation}
%<1.4>
\Delta^2\hat J\Delta^2(\hat\varphi)\ge\frac{\hbar^2}{4},
\end{equation}
where
\begin{equation}
%<1.5>
\Delta^2(\hat\varphi) = \frac{1-|\langle U\rangle|^2}{|\langle
U\rangle|^2},
\end{equation}
where $U=e^{{\rm i}\hat\varphi}$, was made in [10].  Namely, an
upper bond $\hbar$ was found therein for the product of
uncertainties $\Delta\hat J$ and $\Delta(\hat\varphi)$ in the
(normalized) coherent state $ |\xi\rangle$, such that
\begin{equation}
%<1.6>
\hbar > \Delta_\xi\hat J\Delta_\xi(\hat\varphi) > \frac{\hbar}{2}.
\end{equation}
We point out that $\Delta(\hat\varphi)$ cannot be identified with
any uncertainty of the angle.  Indeed, in the eigenvector of $\hat
J$ we have $\Delta\hat J=0$ and $\Delta(\hat\varphi)=\infty$.  But
the maximal uncertainty for the position of a particle on a circle
is $\pi$, so $\Delta(\hat\varphi)$ should be taken modulo $\pi$.
Obviously, $\Delta\hat J=0$ and $\Delta(\hat\varphi)\le\pi$ violate
the inequality (1.4), a contradiction.  It also seems unlikely that
the condition (1.6) allows to determine uniquely the coherent states.
It thus appears that the meaning of (1.4) both in the context of the
quantum mechanics on a circle and corresponding coherent states is dim.
We finally remark that the uncertainty relation (1.4) is implied by
(2.5) and the following inequality [11]:
\begin{equation}
%<1.7>
\langle A^\dagger A+AA^\dagger\rangle\langle B^\dagger B+BB^\dagger 
\rangle \ge |\langle A^\dagger B-BA^\dagger\rangle|^2,
\end{equation}
where we set $A=\hat J-\langle \hat J\rangle$ and $B=U-\langle U
\rangle$.
\section{Coherent states for the quantum mechanics on a circle}
In this section we summarize the elementary facts about the coherent states
for a quantum particle on a circle [9,10].  We begin by recalling
the basic properties of the quantum mechanics on a circle.  Consider a 
free particle on a circle $S^1$.  The classical Hamiltonian is given by
\begin{equation}
%<2.1>
H = \hbox{$\frac{1}{2}$}J^2,
\end{equation}
where $J$ is the angular momentum and we have assumed for simplicity
that the particle has unit mass and it moves in a unit circle.
Clearly, we have the Poissson bracket of the form
\begin{equation}
%<2.2>
\{\varphi,J\} = 1,
\end{equation}
where $\varphi$ is the angle specifying the position on a circle.
The Poisson bracket (2.2) leads according to the rules of the canonical 
quantization to the commutator
\begin{equation}
%<2.3>
[\hat\varphi,\hat J] = {\rm i},
\end{equation}
where we set $\hbar=1$.  It can be demonstrated that the commutator
(2.3) is defined only on the zero vector.  Therefore, the better
candidate than $\hat\varphi$ for representing the position of a
quantum particle on a circle is the unitary operator $U$ such that
\begin{equation}
%<2.4>
U = e^{i\hat\varphi}.
\end{equation}
An immediate consequence of (2.3) and (2.4) is the following algebra:
\begin{equation}
%<2.5>
[\hat J,U] = U.
\end{equation}
We also point out that (2.5) can be obtained directly from (1.1) and (2.4).  Consider 
the eigenvalue equation
\begin{equation}
%<2.6>
\hat J|j\rangle = j|j\rangle.
\end{equation}
From (2.5) and (2.6) it follows that the operators $U$ and
$U^\dagger $ are the ladder operators.  Namely
\begin{equation}
%<2.7>
U|j\rangle = |j+1\rangle,\qquad U^\dagger |j\rangle = |j-1\rangle.
\end{equation}
Demanding the time-reversal invariance of the algebra (2.5) we find [9]
that the eigenvalues $j$ of the operator $\hat J$ can be only
integer or half-integer.  In this work we restrict, for simplicity,
to the case with integer $j$.  We finally write down the orthogonality
and completeness conditions satisfied by the vectors $|j\rangle$ such that
\begin{eqnarray}
%<2.8>
\langle j|k\rangle &=& \delta_{jk},\\
\sum_{j=-\infty}^{\infty} |j\rangle\langle j|&=& I.
\end{eqnarray}

We now collect the basic facts about the coherent states for a
particle on a circle.  These states can be defined by means of
the eigenvalue equation [9]
\begin{equation}
%<2.10>
Z|\xi\rangle = \xi|\xi\rangle,
\end{equation}
where $Z = e^{-\hat J + \hbox{$\frac{1}{2}$}}U$, and the complex
number $\xi = e^{-l + {\rm i}\varphi}$ parametrizes the circular cylinder
which is the classical phase space for the particle moving in a
circle.  We remark that in view of the identity
\begin{equation}
%<2.11>
Z = e^{{\rm i}(\hat\varphi +{\rm i}\hat J)},
\end{equation}
(2.10) has the form analogous to the eigenvalue equation satisfied
by the standard coherent states $|z\rangle$ such that 
\begin{equation}
%<2.12>
e^{ia}|z\rangle = e^{iz}|z\rangle,
\end{equation}
where $a\sim \hat q+i\hat p$ is the standard Bose
annihilation operator and $\hat q$ and $\hat p$ are the
position and momentum observables, respectively.
The projection of the vectors $|\xi\rangle$ onto the basis
vectors $|j\rangle$ is given by
\begin{equation}
%<2.13>
\langle j|\xi\rangle = \xi^{-j}e^{-\frac{j^2}{2}}.
\end{equation}
On using the parameters $l$ and $\varphi$ (2.13) can be written in the
following equivalent form:
\begin{equation}
%<2.14>
\langle j|l,\varphi\rangle =e^{lj-{\rm i}j\varphi}e^{-\frac{j^2}{2}},
\end{equation}
where $|l,\varphi\rangle\equiv|\xi\rangle$ with $\xi = e^{-l + {\rm
i}\varphi}$.  The coherent states are not orthogonal.  We have
\numparts
%<2.15>
\begin{eqnarray}
\langle \xi|\eta\rangle &=&
\sum_{j=-\infty}^{\infty}(\xi^*\eta)^{-j}e^{-j^2} =
\theta_3(\hbox{$\frac{{\rm i}}{2\pi}$}\ln\xi^*\eta|\hbox{$\frac{{\rm i}}
{\pi}$}),\\
\langle l,\varphi|h,\psi\rangle &=&
\theta_3(\hbox{$\frac{1}{2\pi}$}(\varphi-\psi)-\hbox{$\frac{l+h}{2}
\frac{{\rm i}}{\pi}$}|\hbox{$\frac{{\rm i}}{\pi}$}),
\end{eqnarray}
\endnumparts
where $\theta_3$ is the Jacobi theta-function defined by
\begin{equation}
%<2.16>
\theta_3(v|\tau) = \sum_{n=-\infty}^{\infty}q^{n^2}(e^{{\rm i}\pi v})^{2n},
\end{equation}
where $q=e^{{\rm i}\pi\tau}$ and $\hbox{Im}\,\tau>0$.  It follows
immediately from (2.15) that the squared norm of the coherent states
can be written in the form
\numparts
\begin{eqnarray}
%<2.17>
\langle \xi|\xi\rangle &=&
\theta_3(\hbox{$\frac{{\rm i}}{\pi}$}\ln|\xi||\hbox{$\frac{{\rm i}}{\pi}$}),\\
\langle l,\varphi|l,\varphi\rangle &=&
\sum_{j=-\infty}^{\infty}e^{2lj}e^{-j^2}=\theta_3
(\hbox{$\frac{{\rm i}l}{\pi}|\hbox{$\frac{{\rm i}}{\pi}$}$}).
\end{eqnarray}
\endnumparts
The expectation value of the angular momentum $\hat J$ in the
coherent states obeys
\begin{equation}
%<2.18>
\frac{\langle \xi|\hat J|\xi\rangle}{\langle
\xi|\xi\rangle}\approx l,
\end{equation}
where the maximal error arising in the case $l\to0$ is of
order $0.1\%$ and we have the {\em exact\/} equality in the case of
$l$ integer or half-integer.  Therefore, the parameter $l$ labelling
the coherent states can be interpreted as the classical angular
momentum.  The fact that the parameter $\varphi$ can be regarded as
the classical angle is a consequance of the following formula on the
relative expectation value $\langle U\rangle_\xi/\langle U
\rangle_1$ := $\langle\xi|U|\xi\rangle/\langle 1|U |1\rangle$, which
is the most natural candidate to describe the average position on a
circle:
\begin{equation}
%<2.19>
\frac{\langle U\rangle_{\xi}}{\langle U\rangle_1}\approx e^{{\rm i}\phi},
\end{equation}
where the approximation is very good.  More precisely, regardless of the
concrete value of $l$, the maximal error is of order $0.1\%$.  In our opinion
the meaning of (2.18) and (2.19) is that the coherent states are as close as possible 
to the classical phase space.
\section{Uncertainty relations for the quantum mechanics on a circle}
Our purpose now is to introduce the uncertainties of the momentum
and position for a quantum particle on a circle.  We first write
down the following relation implied by (2.9), (2.17), (2.13) and (2.16):
\begin{equation}
%<3.1>
\langle e^{-2\lambda\hat J}\rangle_\xi =\frac{\langle\xi|e^{-2\lambda\hat J}
|\xi\rangle}{\langle\xi|\xi\rangle}=e^{\lambda^2-2l\lambda}\,\frac{\theta_3
(l-\lambda|{\rm i}\pi)}{\theta_3(l|{\rm i}\pi)}.
\end{equation}
On setting $\lambda=\pm1$ in (3.1) we get
\begin{equation}
%<3.2>
\langle e^{-2\hat J}\rangle_\xi=e^{1-2l},\qquad \langle e^{2\hat
J}\rangle_\xi=e^{1+2l}.
\end{equation}
Further, using (2.9), (2.7), (2.13) and (2.16) we find
\begin{equation}
%<3.3>
\langle U^2\rangle_\xi = \frac{\langle\xi|U^2|\xi\rangle}{\langle
\xi|\xi\rangle}=e^{-1}e^{2{\rm i}\varphi}.
\end{equation}
Eqs.\ (3.2) and (3.3) taken together yield the remarkable identity
\begin{equation}
%<3.4>
\langle e^{-2\hat J}\rangle_\xi\langle e^{2\hat J}\rangle_\xi=
\frac{1}{|\langle U^2\rangle_\xi|^2}.
\end{equation}
We now introduce the following measure of the uncertainty of the
angular momentum:
\begin{equation}
%<3.5>
\Delta^2_\phi(\hat J) := \frac{1}{4}\ln\left(\langle e^{-2\hat J}
\rangle_\phi\langle e^{2\hat J}\rangle_\phi\right) 
\end{equation}
and the measure of the uncertainty of the angle
\begin{equation}
%<3.6>
\Delta^2_\phi(\hat \varphi) := \frac{1}{4}\ln\frac{1}
{|\langle U^2\rangle_\phi|^2},
\end{equation}
where $\langle A\rangle_\phi=\langle\phi|A|\phi\rangle/\langle\phi 
|\phi\rangle$, so the identity (3.4) can be written as
\begin{equation}
%<3.7>
\Delta^2_\xi(\hat J)=\Delta^2_\xi(\hat\varphi).
\end{equation}
Notice that both uncertainties (3.5) and (3.6) are nonnegative.  Indeed,
for arbitrary Hermitian operator $X$ we have
\begin{equation}
%<3.8>
\langle e^X\rangle\langle e^{-X}\rangle \ge 1,
\end{equation}
following directly from the Schwarz inequality 
\begin{equation}
%<3.9>
\langle A^\dagger A\rangle\langle B^\dagger B\rangle \ge |\langle
A^\dagger B\rangle|^2,
\end{equation}
by putting $A=e^{\frac{X}{2}}$ and $B=e^{-\frac{X}{2}}$.  An
immediate consequence of (3.8) is nonnegativity of 
$\Delta^2_\phi(\hat J)$.  The inequality
\begin{equation}
%<3.10>
\frac{1}{|\langle U^2\rangle|^2} \ge 1
\end{equation}
ensuring the positivity of $\Delta^2_\phi(\hat \varphi)$ is implied
by the well-known relation
\begin{equation}
%<3.11>
|\langle V\rangle|^2 \le 1,
\end{equation}
which holds true for arbitrary unitary operator $V$.  The inequality
(3.11) can be easily obtained from (3.9) by setting $A=U$ and $B=U^2$.

At first sight the uncertainties (3.5) and (3.6) seem to be weird
without any reference to such measures of uncertainties as a
standard variance.  Nevertheless, we observe that (3.7) has the
identical form as the equation satisfied by the variances of the
momentum and the position in the standard coherent states for a
particle on a real line
\begin{equation}
%<3.12>
\Delta^2_z\hat p = \Delta^2_z\hat q.
\end{equation}
Furthermore, we have the cumulant expansion
\begin{equation}
%<3.13>
\langle e^X\rangle = \exp(\langle\!\langle
X\rangle\!\rangle+\hbox{$\scriptstyle 1\over2!$}\langle\!\langle
X^2\rangle\!\rangle+\hbox{$\scriptstyle 1\over3!$}\langle\!\langle
X^3\rangle\!\rangle+\hbox{$\scriptstyle 1\over4!$}\langle\!\langle
X^4\rangle\!\rangle+\ldots),
\end{equation}
where $\langle\!\langle X^n\rangle\!\rangle$, $n=1$, $2$,
$\ldots\,$, are the cumulants (semiinvariants).  The first four
cumulants are obtained from moments as
\begin{eqnarray}
%<3.14>
\langle\!\langle X\rangle\!\rangle &=& \langle X\rangle,\nonumber\\
\langle\!\langle X^2\rangle\!\rangle &=& \langle X^2\rangle -
\langle X\rangle^2,\nonumber\\
\langle\!\langle X^3\rangle\!\rangle &=& \langle X^3\rangle
-3\langle X^2\rangle\langle X\rangle + 2\langle X\rangle^3,\nonumber\\
\langle\!\langle X^4\rangle\!\rangle &=& \langle X^4\rangle -
4\langle X^3\rangle\langle X\rangle -3\langle X^2\rangle^2
+12\langle X^2\rangle\langle X\rangle^2 -6\langle X\rangle^4.
\end{eqnarray}
Notice that the second cumulant is the usual variance.  The third
and fourth cumulant is called skewness and curtosis, respectively.
Using (3.13), (3.5) and (3.6) we get
\begin{eqnarray}
%<3.15>
\Delta^2_\phi(\hat J) &=&  \langle\!\langle {\hat
J}^2\rangle\!\rangle_\phi
+ \hbox{$\scriptstyle 1\over 3$}\langle\!\langle {\hat
J}^4\rangle\!\rangle_\phi + \hbox{$\scriptstyle 2\over 45$}\langle\!\langle 
{\hat J}^6\rangle\!\rangle_\phi +\ldots ,\\
\Delta^2_\phi(\hat \varphi) &=&  \langle\!\langle {\hat
\varphi}^2\rangle\!\rangle_\phi
- \hbox{$\scriptstyle 1\over 3$}\langle\!\langle {\hat
\varphi}^4\rangle\!\rangle_\phi + \hbox{$\scriptstyle 2\over 45$}\langle\!\langle 
{\hat \varphi}^6\rangle\!\rangle_\phi +\ldots.
\end{eqnarray}
It thus appears that in the first approximation neglecting the
cumulants of order four and higher (even) ones, the uncertainties
(3.5) and (3.6) coincide with the usual variances of the angular
momentum and angle, respectively.  We point out that $\Delta^2_\phi
(\hat J)$ vanishes in the eigenstates $|j\rangle$ of $\hat J$ when
we know the exact value of the angular momentum, and is infinite in
the eigenstate $|\varphi\rangle$ of the operator $U$ corresponding
to the fixed position on a circle.  Analogously, $\Delta^2_\phi(\hat 
\varphi)$ vanishes in the state $|\varphi\rangle$ and is infinite in
the state $|j\rangle$.  We conclude that the uncertainties (3.5) and
(3.6) behave correctly in the states with fixed angular momentum and
angle.  Last but not least we remark that the relations (3.2) and (3.3)
take place also in the case with the half-integer eigenvalues
of $\hat J$.

We now discuss the uncertainty relations for the quantum mechanics
on a circle.  Equations (3.2)--(3.6) taken together yield
\begin{equation}
%<3.17>
\Delta^2_\xi(\hat J)=\hbox{$\scriptstyle 1\over 2$},\qquad \Delta^2_\xi
(\hat\varphi)=\hbox{$\scriptstyle 1\over 2$},
\end{equation}
so
\begin{equation}
%<3.18>
\Delta^2_\xi(\hat J)+\Delta^2_\xi(\hat\varphi)=1.
\end{equation}
The identity (3.18) indicates the following form of the uncertainty
relations for a quantum particle on a circle:
\begin{equation}
%<3.19>
\Delta^2(\hat J)+\Delta^2(\hat\varphi)\ge1,
\end{equation}
minimized at the coherent states.  The uncertainty relations (3.19)
are supported by the numerical calculations (see Fig.\ 1b).  We
finally point out that (3.19) has the form identical as the
uncertainty relations for the sum of variances of the position and
momentum of a particle on a real line implied by the standard
Heisenberg uncertainty relations, of the form
\begin{equation}
%<3.20>
\Delta^2\hat p+\Delta^2\hat q\ge1,
\end{equation}
where we set $\hbar=1$.
\section{Squeezed states for the quantum mechanics on a circle}
We finally study the squeezed states for the quantum mechanics on
a circle and the connected uncertainty relations.  We first observe
that the eigenvectors of the operators $a(s)$ defined as [12]
\begin{equation}
%<4.1>
a(s) = e^{-s\frac{{\hat p}^2}{2}}\hat q e^{s\frac{{\hat p}^2}{2}} =
\hat q + {\rm i}s\hat p,
\end{equation}
where $\hat q$ and $\hat p$ are the standard position and momentum
operators, respectively, and $s>0$ is a real parameter, are the
standard squeezed states.  In analogy to (4.1) we introduce the
operators $Z(s)$ such that
\begin{equation}
%<4.2>
Z(s) = e^{-s\frac{{\hat J}^2}{2}}Ue^{s\frac{{\hat J}^2}{2}}
= e^{-s(\hat J-\frac{1}{2})}U,
\end{equation}
and define the squeezed states $|\xi\rangle_s$ for a quantum
particle on a circle by
\begin{equation}
%<4.3>
Z(s)|\xi\rangle_s = \xi|\xi\rangle_s.
\end{equation}
It should be noted that in view of (4.2) the coherent states for a
particle on a circle satisfying (2.10) correspond to the particular
case $s=1$.  We also remark that we have a generalization of the 
formula (2.11) such that
\begin{equation}
%<4.4>
Z(s) = e^{{\rm i}(\hat\varphi +{\rm i}s\hat J)}.
\end{equation}
Therefore, the squeezed states are related with the scaling of the
angular momentum.  Making use of (4.3), (4.2), (2.6) and (2.7) we easily 
obtain the following generalizations of the relations (2.12) and (2.13):
\numparts
\begin{eqnarray}
%<4.5>
\langle j|\xi\rangle_s &=& \xi^{-j}e^{-\frac{sj^2}{2}},\\
\langle j|l,\varphi\rangle_s &=& e^{lj-{\rm i}j\varphi}e^{-\frac{sj^2}{2}},
\end{eqnarray}
\endnumparts
where $|l,\varphi\rangle_s\equiv|\xi\rangle_s$.  From (4.5) and (2.9)
we derive the overlap integrals such that
\numparts
%<4.6>
\begin{eqnarray}
{}_s\langle \xi|\eta\rangle_s &=&
\sum_{j=-\infty}^{\infty}(\xi^*\eta)^{-j}e^{-sj^2} =
\theta_3(\hbox{$\frac{{\rm i}}{2\pi}$}\ln\xi^*\eta|\hbox{$\frac{{\rm i}s}
{\pi}$}),\\
{}_s\langle l,\varphi|h,\psi\rangle_s &=&
\theta_3(\hbox{$\frac{1}{2\pi}$}(\varphi-\psi)-\hbox{$\frac{l+h}{2}
\frac{{\rm i}}{\pi}$}|\hbox{$\frac{{\rm i}s}{\pi}$}),
\end{eqnarray}
\endnumparts
leading to the following expression on the squared norm of the
squeezed states:
\numparts
\begin{eqnarray}
%<4.7>
{}_s\langle \xi|\xi\rangle_s &=&
\theta_3(\hbox{$\frac{{\rm i}}{\pi}$}\ln|\xi||\hbox{$\frac{{\rm i}s}{\pi}$}),\\
{}_s\langle l,\varphi|l,\varphi\rangle_s &=&
\sum_{j=-\infty}^{\infty}e^{2lj}e^{-sj^2}=\theta_3
(\hbox{$\frac{{\rm i}l}{\pi}|\hbox{$\frac{{\rm i}s}{\pi}$}$}).
\end{eqnarray}
\endnumparts
We point out that the above formulae imply positivity of the
parameter $s$.  We also recall that we study the case of the integer
eigenvalues of the operator $\hat J$.

We now examine the uncertainties in the squeezed states.  Taking
into account (2.9), (2.6), (4.5) and (4.7) we get a generalization of (3.1)
\begin{equation}
%<4.8>
\langle e^{-2\lambda\hat J}\rangle_{\xi,s} =\frac{{}_s\langle\xi|e^{-2
\lambda\hat
J}|\xi\rangle_s}{{}_s\langle\xi|\xi\rangle_s}=e^{\frac{\lambda^2}{s}-
\frac{2l\lambda}{s}}\,\frac{\theta_3
(\frac{l-\lambda}{s}|\frac{{\rm i}\pi)}{s}}{\theta_3(\frac{l}{s}|
\frac{{\rm i}\pi}{s})}.
\end{equation}
Hence, putting $\lambda=\pm s$, we find
\begin{equation}
%<4.9>
\langle e^{-2s\hat J}\rangle_{\xi,s}=e^{s-2l},\qquad \langle e^{2s\hat
J}\rangle_{\xi,s}=e^{s+2l}.
\end{equation}
We have also the generalization of (3.3) of the form
\begin{equation}
%<4.10>
\langle U^2\rangle_{\xi,s} =
\frac{{}_s\langle\xi|U^2|\xi\rangle_s}{{}_s\langle
\xi|\xi\rangle_s}=e^{-s}e^{2{\rm i}\varphi}.
\end{equation}
By (4.9) and (4.10) 
\begin{equation}
%<4.11>
\langle e^{-2s\hat J}\rangle_{\xi,s}\langle e^{2s\hat
J}\rangle_{\xi,s}=
\frac{1}{|\langle U^2\rangle_{\xi,s}|^2},
\end{equation}
which leads to the following most natural generalization
of the uncertainties (3.5) and (3.6) of the angular momentum and
angle, respectively
\begin{eqnarray}
%<4.12>
\widetilde\Delta^2_{\phi,s_0}(\hat J) &=& \frac{1}{4}\ln\left(\langle
e^{-2s_0\hat J}
\rangle_\phi\langle e^{2s_0\hat J}\rangle_\phi\right),\\
\Delta^2_{\phi,s_0}(\hat \varphi) &\equiv& \Delta^2_\phi(\hat \varphi).
\end{eqnarray}
Using the uncertainties (4.12) and (4.13) we arrive at the identity
\begin{equation}
%<4.14>
\widetilde\Delta^2_{\xi,s_0,s_0}(\hat
J)=\Delta^2_{\xi,s_0,s_0}(\hat\varphi)=\hbox{$\scriptstyle s_0\over
2$}
\end{equation}
indicating the generalized uncertainty relations such that
\begin{equation}
%<4.15>
\widetilde\Delta^2_{\phi,s_0}(\hat J)+\Delta^2_\phi(\hat\varphi)\ge s_0,
\end{equation}
where the equality is reached in the squeezed state $|\xi\rangle_{s_0}$.
The uncertainty relations (4.15) are corroborated by the numerical
calculations (see Fig.\ 1 and Fig.\ 2).
\section{Discussion}
In this work we have identified the uncertainties and uncertainty
relations for the quantum mechanics on a circle minimized by the
corresponding coherent states.  We have also introduced the squeezed
states generalizing the coherent states for a quantum particle on a
circle and found the appropriate uncertainty relations saturated by
these states.  Notice that generalized uncertainty relations (4.15),
where the uncertainties are given by (4.12) and (4.13), do not provide 
any criterion for distinguishing coherent and squeezed states as in the case
with the quantum mechanics on a real line.  The situation is even
more complicated in view of the fact that the squeezed states with
different $s$ are not related by a unitary transformation.  Namely,
we have
\begin{displaymath}
 |\xi\rangle_s = e^{-(s-s_0)\hat J^2/2} |\xi\rangle_{s_0}.
\end{displaymath}
Thus the states with different $s$ are not unitarily equivalent and
the problem naturally arises concerning the physical interpretation
of the (dimensionless) parameter $s$.  We point out that in the case
of the standard squeezed states for a particle on a real line the
states with different squeezing are related by a unitary
transformation.

\Figures
\begin{figure}
\caption{The plot of $\widetilde\Delta^2_{\xi,s,s_0}(\hat
J)+\Delta^2_{\xi ,s}(\hat\varphi)$ (compare (4.15)), for a)
$s_0=.5$, b) $s_0=1$, and c) $s_0=1.5$, where $\widetilde
\Delta^2_{\xi,s,s_0}(\hat J) = \frac{1}{4}
\ln\left(\langle e^{-2s_0\hat J}\rangle_\phi\langle e^{2s_0\hat J}
\rangle_\phi\right)$ and the expectation values from the argument of 
the logarithm are determined from (3.1) with $\lambda=\pm s_0$ and
$l=1$; the uncertainty of the angle given by (4.13), (3.6) and (4.10) is $s/2$.  
In accordance with (4.15) and (4.14) the coordinates of the minima are 
$(s_0,s_0)$ (see also Fig.\ 2).  We point out that the case b) with $s_0=1$
refers to the coherent states.  More precisely, we have $\widetilde
\Delta^2_{\xi,s,s_0}(\hat J)+\Delta^2_{\xi ,s}(\hat\varphi)\equiv
\Delta^2_{\xi,s}(\hat J)+\Delta^2_{\xi ,s}(\hat\varphi)$.}
\label{fig1}
%\end{figure}
%\begin{figure}
\caption{The plot of minima of the function from Fig.\ 1.  As
expected in view of (4.15) and (4.14) $s_{\rm min}=s_0$, and $\widetilde
\Delta^2_{\xi,s_{\rm min},s_0}(\hat J)+\Delta^2_{\xi ,s_{\rm min}}
(\hat\varphi)=s_0$.}
\label{fig2}
%\end{figure}
%\begin{figure}
\end{figure}
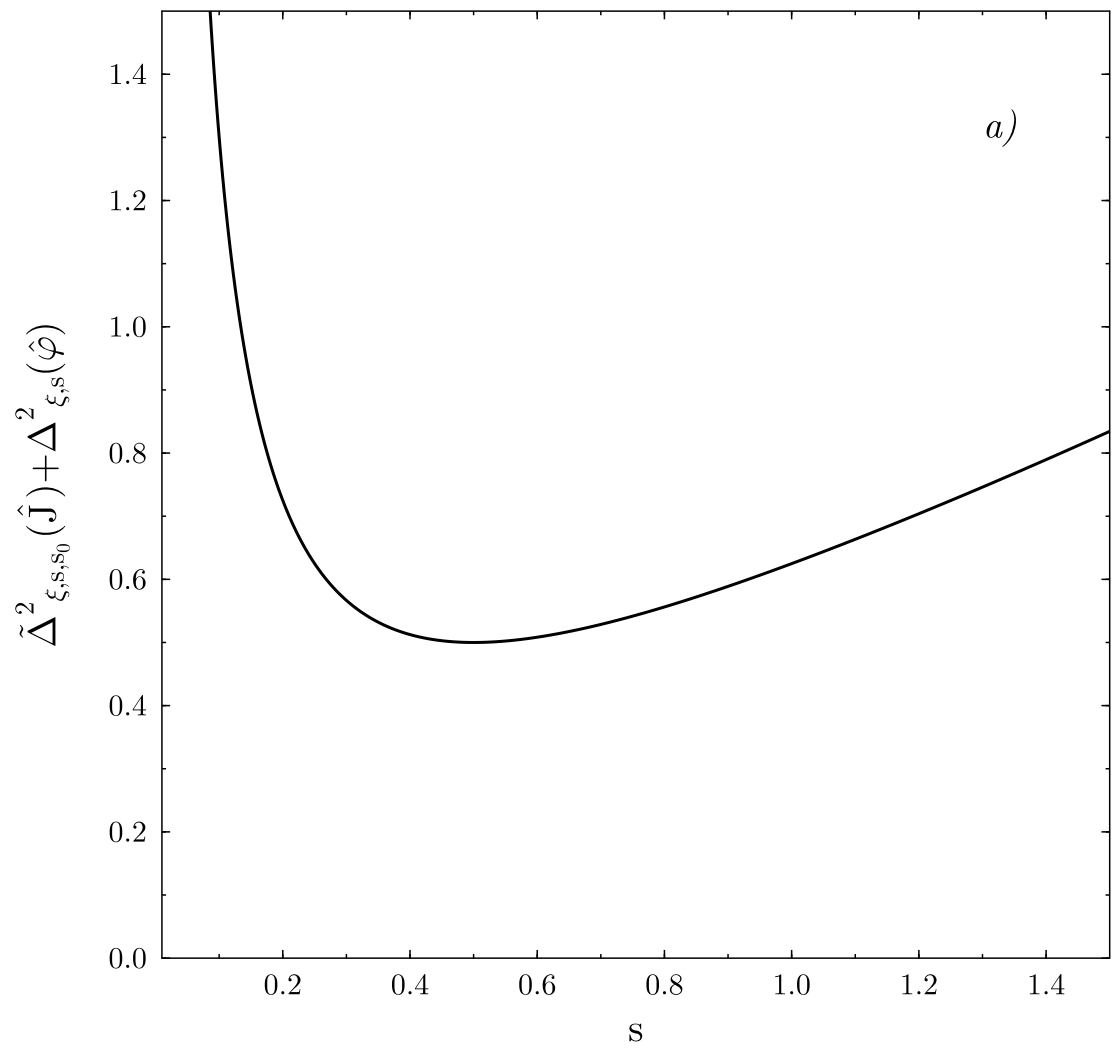
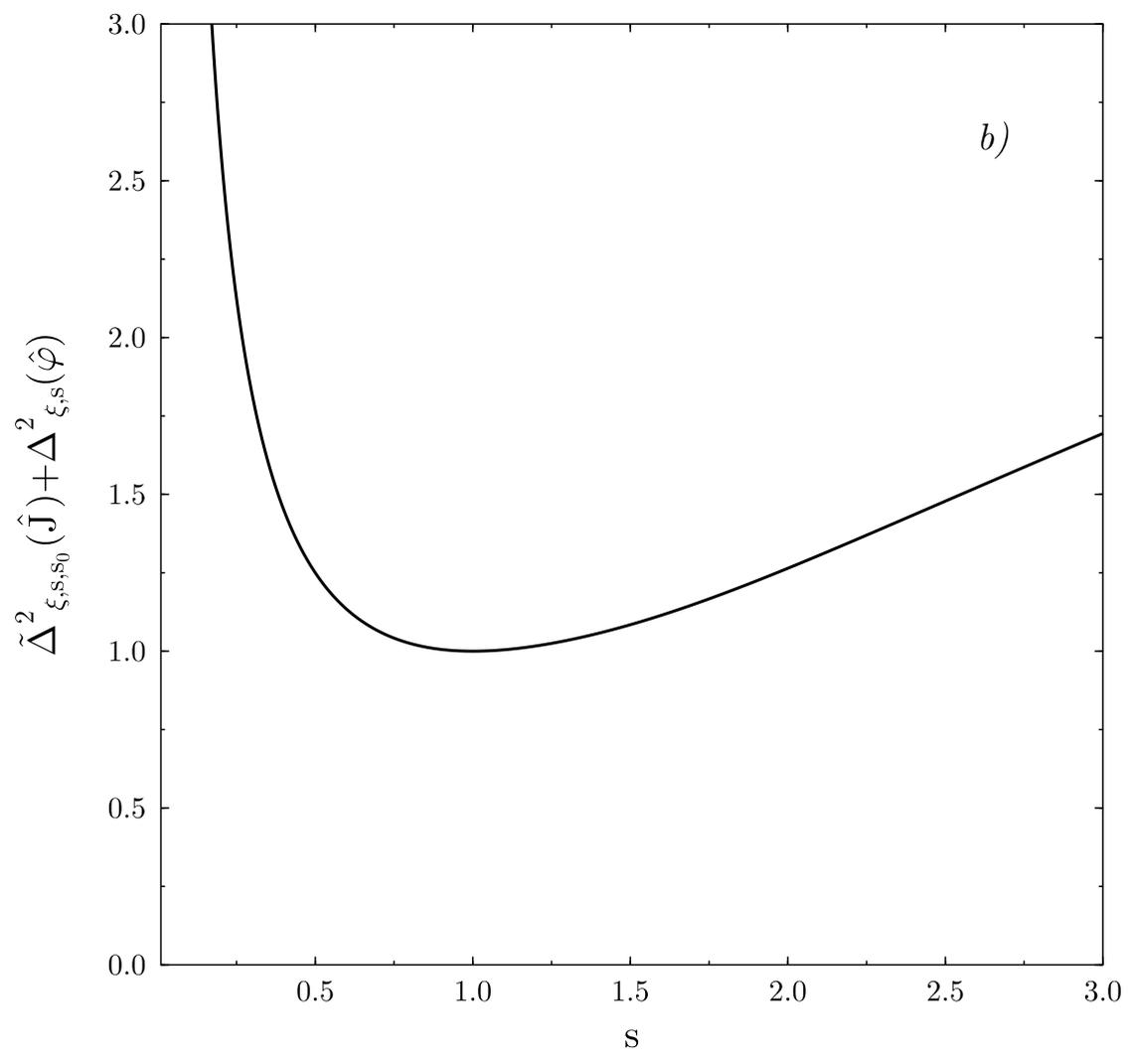
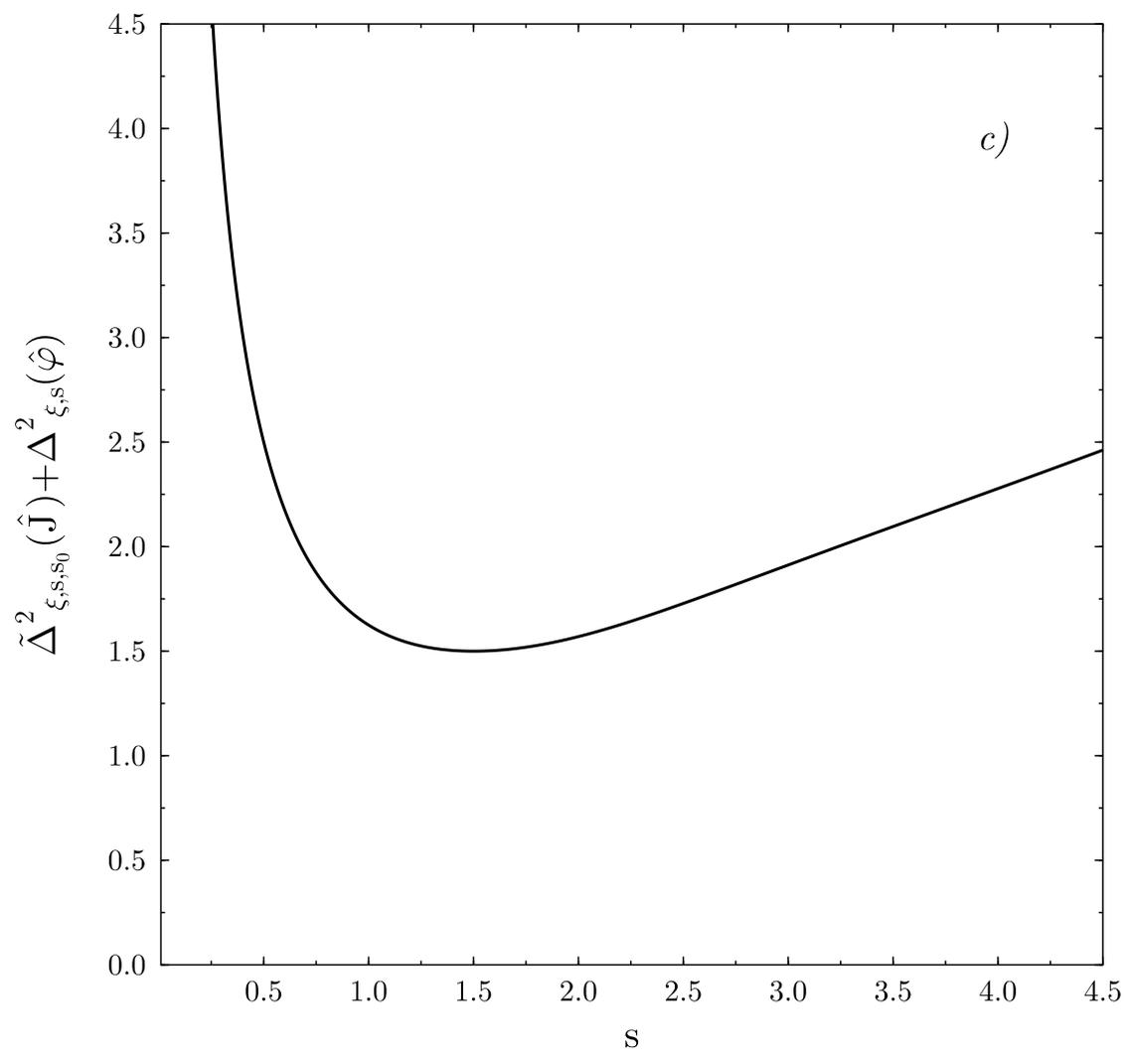
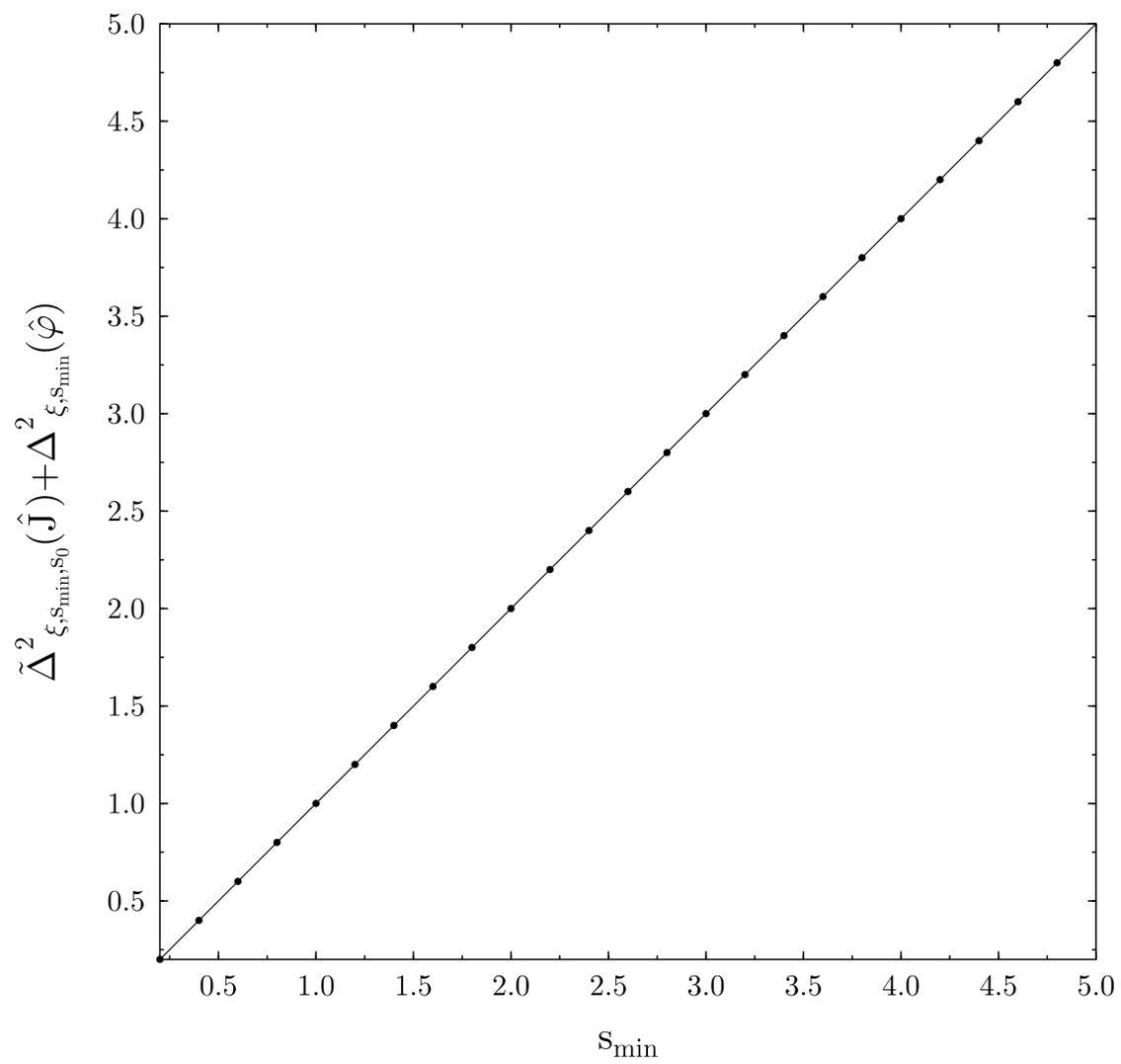
\end{document}